\documentclass[fleqn,usenatbib]{mnras}

\DeclareRobustCommand{\VAN}[3]{#2}
\let\VANthebibliography\thebibliography
\def\thebibliography{\DeclareRobustCommand{\VAN}[3]{##3}\VANthebibliography}

\usepackage{graphicx}	
\usepackage{amsmath}	
\usepackage{amssymb}	
\usepackage{bm}
\usepackage{multirow}

\usepackage[T1]{fontenc}
\usepackage{subcaption}

\newcommand\Eq[1]{Eq.\,(\ref{#1})}
\newcommand\Fig[1]{Fig.~\ref{#1}}
\newcommand\Sec[1]{Sec.~\ref{#1}}
\newcommand\Tab[1]{Tab.~\ref{#1}}

\newcommand{\Mach}{{\mathcal{M}}}      
\newcommand{\Pm}{\text{Pm}}
\renewcommand{\Re}{\text{Re}} 
\newcommand{\Rm}{\text{Rm}}

\renewcommand{\vec}[1]{\boldsymbol{#1}}	
\newcommand{\dd}{\mathrm{d}}        

\newcommand{\G}{{\rm G}}      

\newcommand{\Emag}{E_{\rm{m}}/E_{\rm{m0}}}

\newcommand{\ratio}{E_{\rm{mag}}/E_{\rm{kin}}}
\newcommand{\Rmcrit}{{\rm Rm}_{\rm crit}}

\newcommand{\Gammasat}{\Gamma_{\rm sat}}
\newcommand{\expcoeff}{\alpha}

\newcommand{\magnetisation}{r_{\rm Larmor}/L_{\rm box}}
\newcommand{\initmagnetisation}{r_{\rm Larmor,0}/L_{\rm box}}
\newcommand{\Einit}{(E_{\rm{mag}}/E_{\rm{kin}})_{0}}
\newcommand{\betainit}{\beta_{0}}
\newcommand{\ngrid}{N_{\rm{grid}}}
\newcommand{\nppc}{N_{\rm{ppc}}}

\newcommand{\vel}{\vec{v}}
\newcommand{\pos}{\vec{r}}
\renewcommand{\vec}[1]{\mathbf{#1}}
\newcommand{\Vturb}{v_{\rm turb}}
\newcommand{\Vtherm}{v_{\rm th}}
\newcommand{\Lturb}{L_{\rm turb}}
\newcommand{\tth}{t_{\rm th}}
\newcommand{\tcool}{t_{\rm cool}}
\newcommand{\ted}{t_{\rm turb}}

\newcommand{\E}{\vec{E}}
\newcommand{\B}{\vec{B}}
\newcommand{\J}{\vec{J}}

\newcommand{\ahkash}{\texttt{AHKASH}\,}
\newcommand{\flash}{\texttt{FLASH}\,}

\title[Supersonic weakly-collisional turbulent dynamo]{Critical magnetic Reynolds number of the weakly-collisional turbulent dynamo in supersonic plasmas}

\author[Achikanath Chirakkara et al.]{
Radhika Achikanath Chirakkara,$^{1,2}$\thanks{E-mail: \href{mailto:rchirakkara@cita.utoronto.ca}{rchirakkara@cita.utoronto.ca}}
and Christoph Federrath,$^{2}$\thanks{E-mail: \href{christoph.federrath@anu.edu.au}{christoph.federrath@anu.edu.au}}
\\
$^{1}$Canadian Institute for Theoretical Astrophysics, University of Toronto, Toronto, M5S 3H8, ON, Canada\\
$^{2}$Research School of Astronomy and Astrophysics, Australian National University, Canberra, ACT 2611, Australia
}

\date{Accepted XXX. Received YYY; in original form ZZZ}

\pubyear{\the\year{}}

\begin{document}
\label{firstpage}
\pagerange{\pageref{firstpage}--\pageref{lastpage}}
\maketitle

\begin{abstract}
Weakly-collisional plasmas are found in several turbulent astrophysical environments, such as the Intra-galaxy-cluster medium (ICM) and the solar wind, where magnetic fields can be amplified by the turbulent dynamo. The onset of dynamo action is determined by the critical magnetic Reynolds number, $\Rmcrit$, above which turbulent motions amplify magnetic energy. Although $\Rmcrit$ has been studied extensively in collisional magnetohydrodynamic (MHD) turbulence, its value in supersonic weakly-collisional plasmas remains unknown. Here we determine $\Rmcrit$ using three-dimensional hybrid particle-in-cell (HPIC) simulations of solenoidally driven turbulence with a sonic Mach number $\Mach\approx2$. We vary the magnetic Reynolds number over $\Rm\approx31$--$1000$ and consider two initial plasma magnetisations, characterised by initial Larmor ratios $r_{\rm Larmor, 0}/L_{\rm box}=1$ and $100$. We find that the magnetic energy grows exponentially at sufficiently large $\Rm$ and decays when magnetic diffusion dominates at low $\Rm$. By fitting the magnetic-energy growth rate as a function of $\Rm$, we measure $\Rmcrit=190^{+21}_{-17}$ and $198^{+16}_{-18}$ for the initially magnetised and unmagnetised plasma, respectively. These values are the same within their uncertainties, indicating that the initial plasma magnetisation does not significantly affect $\Rmcrit$ in the supersonic regime. The measured values are higher than those previously obtained for the subsonic weakly-collisional dynamo, showing that increasing compressibility makes weakly-collisional turbulent dynamo action more difficult to excite. The measured values are also approximately 1.5 times higher than the $\Rmcrit$ value reported for supersonic MHD turbulence.
\end{abstract}

\begin{keywords}
dynamo -- turbulence --  magnetic fields -- methods: numerical -- galaxies: clusters: intracluster medium -- plasmas 
\end{keywords}



\section{Introduction}
\label{sec:introduction}
Weakly-collisional plasmas are found in a broad range of astrophysical environments, including the solar wind, the hot intracluster medium (ICM) of galaxy clusters, hot accretion flows, and collisionless shocks. In these systems, the Coulomb mean free path is comparable to the characteristic macroscopic length scales. Therefore, a collisional magnetohydrodynamic (MHD) description does not correctly capture the plasma evolution, and a kinetic treatment is required \citep{Kulsrud2005,Schekochihin&Cowley2006,Simionescu+2019,Kunz+2022}. The weak collisionality of these plasmas causes departures from local thermodynamic equilibrium and gives rise to kinetic instabilities that regulate the transport of momentum and energy.

Many weakly-collisional astrophysical plasmas are also turbulent. In galaxy clusters, turbulence can be generated by merger activity, and energy injection from active galactic nuclei
\citep{Subramanian+2006,Banerjee&Sharma2014,Mohapatra&Sharma2019, Zhuravleva+2014, Zhuravlevaetal2019, Gatuzz+2023}. The compressibility of these motions varies substantially among different environments. Turbulence can be subsonic in relaxed cluster interiors \citep{HitomiCollaboration2016,Gatuzz+2022a,Gatuzz+2022b,Gatuzz+2023}, whereas merging systems and shocked regions can exhibit transonic or supersonic motions \citep{Hoang+2017}.

Magnetic fields are dynamically important in many of these systems. For example, magnetic fields of order ${\sim}\mu\G$ are observed in the hot ICM \citep{Bonafedeetal2010,Bonafede+2013,Botteon+2022}. These strong fields can grow from much weaker primordial or astrophysical seed fields through turbulent dynamo action, which converts the kinetic energy of turbulent motions into magnetic energy \citep{Kazantsev1968,Moffatt1978,B&S2005}. During the kinematic stage of the dynamo, turbulent motions stretch, twist, and fold the weak magnetic field, causing exponential growth of the magnetic energy. The subsequent back-reaction of the magnetic field on the flow eventually leads to the saturated state of the dynamo \citep{CFetal11, Seta&Federrath2020, Seta&Federrath2021b, AchikanathEtAl2021}.

Whether a turbulent flow can amplify magnetic energy is determined by the competition between inductive amplification and magnetic diffusion. This competition is quantified by the magnetic Reynolds number, $\Rm$, defined as
\begin{equation}
    \Rm = \frac{\Vturb\Lturb}{\eta},
\end{equation}
where $\Vturb$, $\Lturb$, and $\eta$ are the turbulent speed, turbulence driving scale, and Ohmic diffusivity, respectively.
Dynamo action is possible only when $\Rm$ exceeds a critical magnetic Reynolds number, $\Rmcrit$. For $\Rm<\Rmcrit$, magnetic diffusion dissipates magnetic energy more rapidly than turbulent motions can generate it. The value of $\Rmcrit$ therefore provides an important measure of the conditions required to excite the turbulent dynamo.

Studies of the collisional MHD dynamo have demonstrated that $\Rmcrit$ is not universal. It depends on the magnetic Prandtl number, $\Pm$, as well as on the compressibility of the plasma \citep{Haugen+2004b,Schoberetal2012,Federrathetal2014ApJ,Seta+2020}. At fixed $\Pm$, numerical simulations find that the critical magnetic Reynolds number increases as the turbulence transitions from the subsonic to the supersonic regime. For $\Pm=1$, for example, $\Rmcrit$ increases from approximately 35 in subsonic flows to 70 in supersonic flows \citep{Haugen+2004b}. This increase is likely associated with the formation of shocks and strong compressive structures, as dynamo amplification is driven primarily by solenoidal motions, which efficiently stretch, twist, and fold magnetic-field lines, whereas compressive motions contribute less efficiently. In highly compressible, supersonic simulations, \citet{Federrathetal2014ApJ} measured $\Rmcrit=129^{+43}_{-31}$. Together, these results suggest that increasing compressibility reduces the efficiency of magnetic-field amplification and raises the threshold for dynamo action.

Analytical studies based on the Kazantsev model also predict that the threshold for dynamo action depends strongly on the nature of the turbulence. In particular, the predicted $\Rmcrit$ is considerably larger for highly compressible, Burgers-like turbulence than for incompressible, Kolmogorov-like turbulence \citep{Schoberetal2012,Bovino+2013}. Both analytical and numerical studies indicate that compressibility can inhibit the onset of the turbulent dynamo. These results have been obtained for collisional MHD plasmas, and it remains unclear whether the same behaviour applies to weakly-collisional plasmas.

The turbulent dynamo in weakly-collisional plasmas has only recently been studied using kinetic and hybrid-kinetic numerical methods \citep{Rinconetal2016,St-Onge&Kunz2018,St-Ongeetal2020,Sironi+2023,Zhou+2024,AchikanathChirakkara+2023,AchikanathChirakkara+2024, AchikanathChirakkara+2025}. In the hybrid particle-in-cell (HPIC) approach, ions are evolved kinetically as charged particles, while electrons are represented as a fluid. In collisional MHD simulations, the viscosity and Ohmic diffusivity can be set independently, allowing the kinetic Reynolds number, $\Re$, and the magnetic Prandtl number, $\Pm=\Rm/\Re$, to be controlled separately. In a weakly-collisional HPIC plasma, however, the effective viscosity arises self-consistently from wave--particle interactions and may evolve as the magnetic field and plasma magnetisation change \citep{Schekochihin&Cowley2006}. The corresponding $\Re$ and $\Pm$ are therefore emergent properties rather than fixed input parameters. This complicates direct comparisons between HPIC and MHD dynamos.

Using HPIC simulations, \citet{AchikanathChirakkara+2023} measured the critical magnetic Reynolds number of the weakly-collisional turbulent dynamo in the subsonic regime. However, the critical magnetic Reynolds number of the weakly-collisional dynamo has not been determined in the supersonic regime. In this work, we investigate the onset of weakly-collisional turbulent dynamo action in a supersonic plasma.

The rest of this paper is organised as follows. We describe the hybrid-kinetic equations, plasma cooling, turbulent driving, and initial conditions in \Sec{sec:methods}. In \Sec{sec:results}, we examine the density, velocity, and magnetic-field structures, follow the time evolution of the magnetic energy, and determine the critical magnetic Reynolds number. We compare our results with previous studies of collisional MHD and subsonic weakly-collisional plasmas in \Sec{sec:comparison}. Finally, we present our conclusions in \Sec{sec:conclusions}.

\section{Methods}
\label{sec:methods}
We perform numerical experiments of the weakly-collisional dynamo using the HPIC code \ahkash \citep{AchikanathChirakkara+2024}. In this section, we describe the hybrid-kinetic equations solved by \ahkash, the cooling prescription used to maintain an isothermal plasma, and the external
forcing used to drive turbulence. We then discuss the plasma parameters and the initial conditions of our simulations.

\subsection{Hybrid-kinetic equations}
\label{sec:hybridkinetic_eqns} Within the hybrid-kinetic framework, ions are evolved kinetically as particles, while the electrons are represented as a massless fluid. The evolution of the $p^{\rm th}$ ion is governed by the equations of motion,
\begin{align}
        & \frac{\dd \pos_{p}} {\dd t} = \vel_{p},
        \label{eqn:part_evolution1} \\
        & \frac{\dd \vel_{p}}{\dd t}
        = \frac{q_{\rm i}}{m_{\rm i}}
        \left(\E+\vel_{p}\times\B\right)
        +\vec{f},
        \label{eqn:part_evolution2}
\end{align}
where $\pos_p$ and $\vel_p$ denote the particle position and velocity, respectively, with $p=1,\ldots,P$ and $P$ the total number of particles. Here, $q_{\rm i}$ and $m_{\rm i}$ are the ion charge and mass. The particle trajectories are determined by the Lorentz force together with the turbulence forcing term, $\vec{f}$, which injects turbulent kinetic energy into the plasma (Section~\ref{sec:turb_driving}). The evolving particle distribution is used to evaluate the ion moments required to solve for the electromagnetic fields, $\E$ and $\B$. The electric field is obtained from Ohm's law \citep{Rosin+2011},
\begin{equation}
\E =
\frac{\left(\J-\J_{\rm i}\right)\times\B}{\rho_{\rm i}}
-\frac{\nabla p_{\rm e}}{\rho_{\rm i}}
+\mu_0\eta\J
-\mu_0\eta_{\rm hyper}\nabla^2\J,
\label{eqn:ohms_law}
\end{equation}
where $\J=(\nabla\times\B)/\mu_0$ is the total current density and $\mu_0$ is the magnetic permeability of free space. The ion charge density, $\rho_{\rm i}$, and ion current density, $\J_{\rm i}$, are sampled directly from the particle distribution. We assume an isothermal electron fluid, with the pressure given by the ideal gas equation of state,
\begin{equation}
p_{\rm e}=n_{\rm e}k_{\rm B}T_{\rm e},
\end{equation}
where $n_{\rm e}$ is the electron number density, $T_{\rm e}$ is the constant electron temperature, and $k_{\rm B}$ is Boltzmann's constant. Quasi-neutrality is enforced by the condition, $\rho_{\rm e}=-\rho_{\rm i}$. In Equation~(\ref{eqn:ohms_law}), the coefficient $\eta$ is the Ohmic diffusivity, while the hyper resistivity term, controlled by $\eta_{\rm hyper}$, is included to damp numerical fluctuations at the grid scale. The magnetic field evolves according to Faraday's law,
\begin{equation}
\frac{\partial\B}{\partial t}
=
-\nabla\times\E,
\label{eqn:faradays_law}
\end{equation}
which is solved using the constrained transport algorithm to maintain the divergence-free condition for magnetic fields, $\nabla\cdot\B=0$. Interpolation between particles and the computational grid is performed using the cloud-in-cell method. The ion moments are smoothed to suppress particle noise associated with the finite sampling of the ion distribution function. Additional details of the numerical implementation, together with validation tests of the code, are presented in \citet{AchikanathChirakkara+2024}.

\subsubsection{Cooling for hybrid particle-in-cell simulations}
\label{sec:cooling}
Turbulent dissipation and shocks in the supersonic regime continuously heat the plasma. To maintain a statistically steady, isothermal state, this heating must be balanced by an appropriate cooling mechanism. While the electron fluid is kept isothermal using an isothermal equation of state (Section~\ref{sec:hybridkinetic_eqns}), no corresponding cooling mechanism exists for the kinetic ions. Therefore, an explicit cooling procedure is required to regulate the ion temperature and maintain thermal equilibrium between the ions and electrons.

To achieve this, we use the local ion-cooling scheme introduced by \citet[see their sec.~2.11]{AchikanathChirakkara+2024}, which rescales the ion thermal speed, $\Vtherm$. Cooling is applied on a timescale of $(\Delta t)_{\rm cool}=0.01\,\tcool$, where $\tcool=\tth/\Mach$ and $\tth=\Delta l/\Vtherm$ is the thermal crossing time, defined as the time required for a sound wave to travel across a grid cell of size $\Delta l$. The sonic Mach number,
\begin{equation}
    \Mach = \frac{\Vturb}{\Vtherm},
\end{equation}
where $\Vturb$ and $\Vtherm$ are the turbulent and thermal speeds, respectively.
This cooling prescription maintains an approximately constant plasma temperature in space and time throughout the simulation.

\subsection{Turbulent driving}
\label{sec:turb_driving}
The external turbulent forcing, $\vec{f}$, is generated using the \texttt{TurbGen} method \citep{Federrath+2010,FederrathEtAl2022ascl}, which uses an Ornstein--Uhlenbeck stochastic process. We use purely solenoidal forcing, $\nabla\cdot\vec{f}=0$ here and leave an investigation of fully compressive forcing to future work. Energy is injected only on large spatial scales, corresponding to wavenumbers in the range $k_{\rm turb}L_{\rm box}/(2\pi)\in[1,3]$, where $L_{\rm box}$ is the side length of the computational box. The forcing spectrum follows a parabolic function that has its maximum at $k_{\rm turb}L_{\rm box}/(2\pi)=2$ and decreases smoothly to zero at the boundaries of the driving range, $k_{\rm turb}L_{\rm box}/(2\pi)=1$ and $3$. This is standard procedure in turbulence simulations \citep{FederrathOffner2026}. The turbulent crossing time is defined as 
\begin{equation}
    \ted = \frac{\Lturb}{\Vturb},
\end{equation}
where $\Lturb=L_{\rm box}/2$ is the turbulent driving scale. 

\subsection{Plasma parameters and initial conditions}
 \label{sec:pars_ics}
This section describes the physical parameters and initial conditions used in the HPIC runs. All simulations are performed in a three-dimensional periodic domain with a grid resolution of $\ngrid^3=128^3$ cells and $\nppc=100$ particles per cell. The convergence tests for the grid and particle resolutions are provided in our previous studies \citep{AchikanathChirakkara+2024, AchikanathChirakkara+2025}.

\subsubsection{Plasma parameters}
\label{sec:pars}
The properties of the turbulent dynamo depend on the following plasma parameters. The compressibility of the plasma is characterised by the Mach number. In this study, we focus on the supersonic regime, with $\Mach = 2$. We consider six magnetic Reynolds numbers, $\Rm = 31.3,\ 62.5,\ 125,\ 250,\ 500,$ and $1000$, where $\Rm$ is controlled by varying the Ohmic diffusivity $\eta$. To suppress numerical fluctuations at the grid scale, we include a hyper-resistive term (cf.~Eq.~\ref{eqn:ohms_law}). The corresponding hyper-resistive Reynolds number is defined as
\begin{equation}
    \Rm_{\rm hyper} =
    \frac{\Vturb\Lturb^3}{\eta_{\rm hyper}}.
\end{equation}
We use $\Rm_{\rm hyper}=\ngrid^{10/3}$, where $\ngrid$ is the grid resolution (see sec.~2.5.2 in \citealt{AchikanathChirakkara+2023}).
In weakly-collisional plasmas, the effective viscosity arises self-consistently through wave--particle interactions. As a result, both the kinetic Reynolds number, $\Re$, and the magnetic Prandtl number, $\Pm=\Rm/\Re$, are emergent properties of the plasma rather than prescribed parameters in our simulations.

\subsubsection{Initial conditions and parameter study}
\label{sec:ics_table}
We initialise each simulation with a weak magnetic field. The initial magnetic-to-turbulent kinetic energy ratio is defined as
\begin{equation}
    \Einit=
    \frac{B_0^{2}/(2\mu_{0})}
         {\rho_{\rm m}\Vturb^{2}/2} = \frac{2}{\betainit \Mach^{2}},
\end{equation}
where $B_0$ is the initial magnetic field strength and $\rho_{\rm m}$ is the ion mass density. The initial plasma beta is denoted by $\betainit$. Throughout this work, we set $\Einit=10^{-10}$ to allow for turbulent dynamo amplification in the kinematic regime.

The level of plasma magnetisation is characterised by the Larmor ratio, defined as the ratio of the ion Larmor radius to the size of the computational domain \citep{AchikanathChirakkara+2023}. The initial Larmor ratio is given by
\begin{equation}
    \frac{r_{\rm Larmor, 0}}{L_{\rm box}} =
    \frac{m_{\rm i}\Vtherm}
         {q_{\rm i}B_0L_{\rm box}}.
\end{equation}
We consider two initial magnetisation states, corresponding to $\initmagnetisation=1$ (initially magnetised) and $\initmagnetisation=100$ (initially unmagnetised). More details about the initial conditions and the plasma parameters are provided in sec.~2.5 of \citet{AchikanathChirakkara+2023}. All simulations are performed using the hybrid-precision method, which substantially reduces the computational cost while maintaining numerical accuracy \citep{Federrath+2021,AchikanathChirakkara+2024,FederrathOffner2026}.

The plasma parameters and measured properties of all simulations are summarised in \Tab{table:sims}, including the grid and particle resolutions, Mach number, initial magnetic to kinetic energy ratio, initial plasma beta, initial magnetisation, magnetic Reynolds number, and the resulting growth/decay rate of the magnetic field.

\begin{table*}
\centering
\caption{List of simulations with the corresponding model name, grid resolution ($\ngrid^{3}$), particle-per-cell count ($\nppc$), Mach number ($\mathcal{M}$), initial magnetic to kinetic energy ratio ($\Einit$), initial plasma beta ($\betainit$), initial Larmor ratio ($\initmagnetisation$), magnetic Reynolds number ($\Rm$), and the growth/decay rate of the magnetic energy ($\Gamma$). }
\setlength{\tabcolsep}{2.5pt}
\begin{tabular}{llccccccccc}
\hline
Ser. No & Model & $\ngrid^{3}$ & $\nppc$ & $\mathcal{M}$ & $\Einit$ & $\betainit$ & $\initmagnetisation$ & $\Rm$ & $\Gamma$ ($\ted^{-1}$)\\
\hline

1 & \texttt{Rm1000rL1} & $128^{3}$ & 100 & 2.16$\pm$0.15 & $10^{-10}$ & $5 \times 10^{9}$  & $1$ & 1080$\pm$70 & $\phantom{-}$0.40$\pm$0.03\\

2 & \texttt{Rm500rL1} & $128^{3}$ & 100 & 2.13$\pm$0.14 & $10^{-10}$ & $5 \times 10^{9}$  & $1$ & 532$\pm$34 & $\phantom{-}$0.24$\pm$0.01\\

3 & \texttt{Rm250rL1} & $128^{3}$ & 100 & 2.08$\pm$0.12 & $10^{-10}$ & $5 \times 10^{9}$  & $1$ & 260$\pm$15 & $\phantom{-}$0.11$\pm$0.01\\

4 & \texttt{Rm125rL1} & $128^{3}$ & 100 & 2.06$\pm$0.11 & $10^{-10}$ & $5 \times 10^{9}$  & $1$ & 129$\pm$7 & $-$0.14$\pm$0.07\\

5 & \texttt{Rm63rL1} & $128^{3}$ & 100 & 2.07$\pm$0.11 & $10^{-10}$ & $5 \times 10^{9}$  & $1$ & 64.5$\pm$3.3 & $-$0.61$\pm$0.09\\

6 & \texttt{Rm31rL1} & $128^{3}$ & 100 & 2.08$\pm$0.11 & $10^{-10}$ & $5 \times 10^{9}$  & $1$ & 32.4$\pm$1.8 & $-$1.28$\pm$0.13\\

7 & \texttt{Rm1000rL100} & $128^{3}$ & 100 & 2.11$\pm$0.13 & $10^{-10}$ & $5 \times 10^{9}$  & $10^{2}$ & 1050$\pm$70 & $\phantom{-}$$0.48\pm$0.02\\

8 & \texttt{Rm500rL100} & $128^{3}$ & 100 & 2.07$\pm$0.12 & $10^{-10}$ & $5 \times 10^{9}$  & $10^{2}$ & 518$\pm$30 & $\phantom{-}$0.33$\pm$0.02
\\

9 & \texttt{Rm250rL100} & $128^{3}$ & 100 & 2.06$\pm$0.11 & $10^{-10}$ & $5 \times 10^{9}$  & $10^{2}$ & 257$\pm$13 & $\phantom{-}$0.10$\pm$0.04\\

10 & \texttt{Rm125rL100} & $128^{3}$ & 100 & 2.06$\pm0.10$ & $10^{-10}$ & $5 \times 10^{9}$  & $10^{2}$ & 129$\pm$6 & $-$0.21$\pm$0.10\\

11 & \texttt{Rm63rL100} & $128^{3}$ & 100 & 2.06$\pm$0.12 & $10^{-10}$ & $5 \times 10^{9}$  & $10^{2}$ & 64.2$\pm$3.8 & $-$0.62$\pm$0.08\\

12 & \texttt{Rm31rL100} & $128^{3}$ & 100 & 2.07$\pm$0.11 & $10^{-10}$ & $5 \times 10^{9}$  & $10^{2}$ & 32.3$\pm$1.7 & $-$1.14$\pm$0.12\\
\hline

\end{tabular}
\label{table:sims}
\end{table*}

\section{Results}
\label{sec:results}
We first examine the density, velocity, and
magnetic-field structures in \Sec{sec:slice_plots}, before investigating the time evolution of the magnetic energy and its growth rate during the kinematic regime in \Sec{sec:time_evol}. Finally, in \Sec{sec:critical_Rm} we measure the critical magnetic Reynolds number for turbulent dynamo action in supersonic weakly-collisional plasma.

\subsection{Density, velocity, and magnetic field structures}
\label{sec:slice_plots}
\begin{figure*}
    \centering
    \includegraphics[scale=0.9]{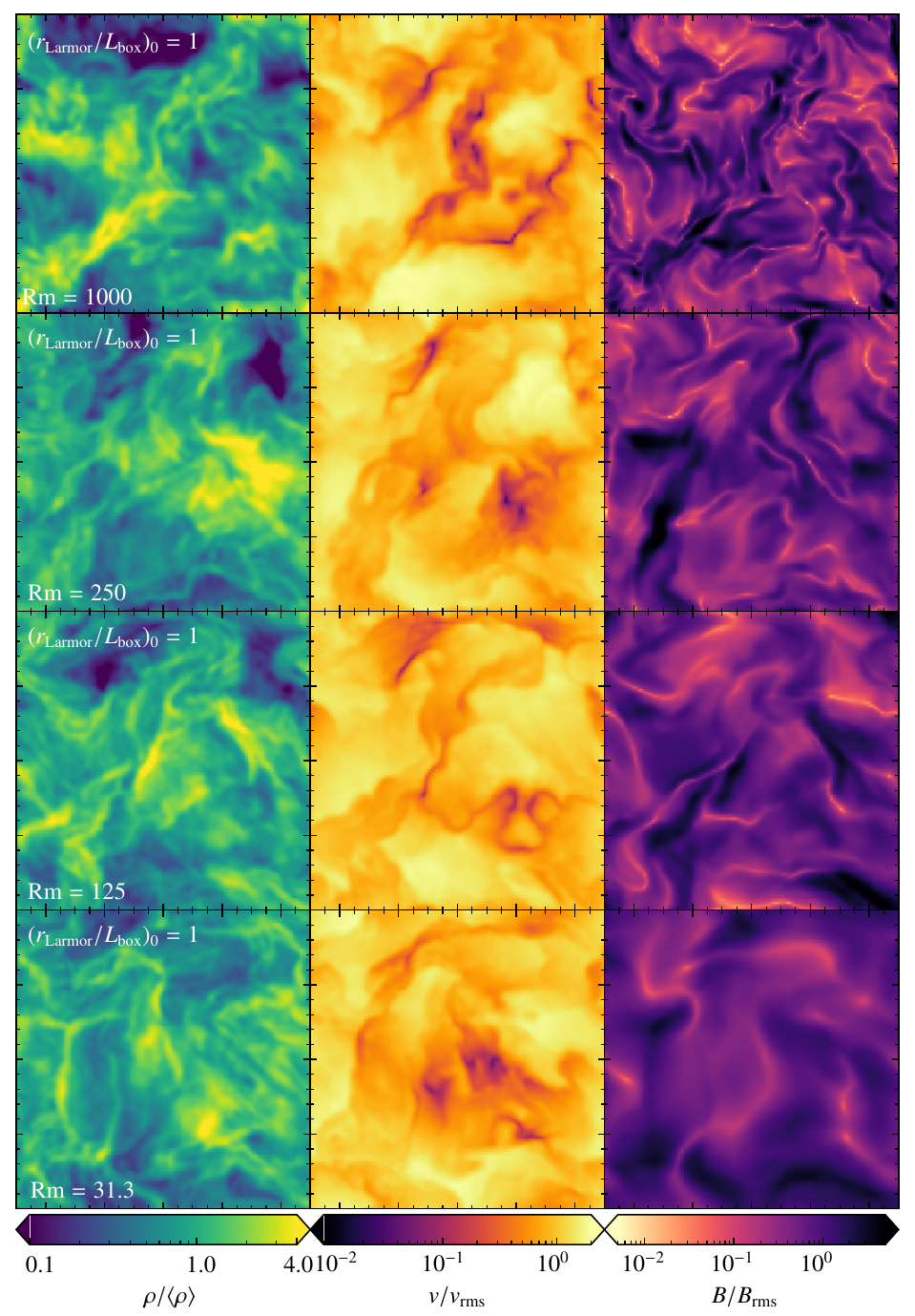}
    \caption{Slice plots of the mass density, normalised to the mean mass density, $\rho/\langle\rho\rangle$ (first column), the velocity magnitude, normalised to the root-mean-square velocity, $v/v_{\rm rms}$ (second column), and the magnetic-field strength, normalised to the root-mean-square magnetic-field strength, $B/B_{\rm rms}$ (third column), for the $\Mach = 2$ HPIC dynamo simulations with $\initmagnetisation = 1$ during the kinematic regime ($t = 25\,\ted$). The density and velocity fields are relatively insensitive to \Rm, whereas the magnetic field depends strongly on it. At lower \Rm, increased diffusivity smooths the magnetic field and suppresses its small-scale structure.}
    \label{fig:slice_plots_rL1}
\end{figure*}

\Fig{fig:slice_plots_rL1} shows slice plots of the density, velocity,
and magnetic fields in the supersonic HPIC simulations during the
kinematic phase at $t=25\,\ted$. From top to bottom, the rows
correspond to $\Rm=1000$, $250$, $125$, and $31.3$. The
corresponding slice plots for an initially unmagnetised plasma with
$\initmagnetisation=100$ are shown in
\Fig{fig:slice_plots_rL100}.

The first column shows the normalised density field. Strong density fluctuations, shocks, and large-scale filamentary structures are present in all simulations, as expected for supersonic turbulence. The density structures are qualitatively similar over the range of magnetic Reynolds numbers studied here, showing that the large-scale supersonic flow is largely unaffected by magnetic diffusivity. The second column of \Fig{fig:slice_plots_rL1} shows the normalised velocity magnitude. The flow is dominated by motions on large scales, consistent with the forcing described in \Sec{sec:turb_driving}. Similar to the density field, the velocity structures exhibit little dependence on the magnetic Reynolds number.

This is expected because we only study the kinematic regime of the dynamo, in which the magnetic field is dynamically weak compared to the velocity field and its back-reaction on the flow is negligible. Varying $\Rm$ therefore affects the magnetic-field evolution without changing the density or velocity fields in the kinematic regime. Differences may emerge in the nonlinear and saturation regimes of the dynamo as the magnetic field strength increases and becomes comparable to the velocity field. The filamentary density structures are spatially correlated
with the velocity field, consistent with previous studies of
supersonic turbulence
\citep{Padoan+1997,Kowal+2007,Kritsuk+2007,Federrath+2010}.

The third column of \Fig{fig:slice_plots_rL1} shows the normalised magnetic-field strength. Unlike the density and velocity fields, the magnetic-field morphology depends strongly on $\Rm$. At large $\Rm$, the field exhibits intermittent structure over small spatial scales. These features become smoother as $\Rm$ decreases, because higher
Ohmic diffusion removes small-scale magnetic structure. 

\subsection{Time evolution}
\label{sec:time_evol}
\begin{figure*}
\begin{center}
\def\arraystretch{0}
\setlength{\tabcolsep}{5pt}
\begin{tabular}{ll}
\includegraphics[height=0.65\linewidth]{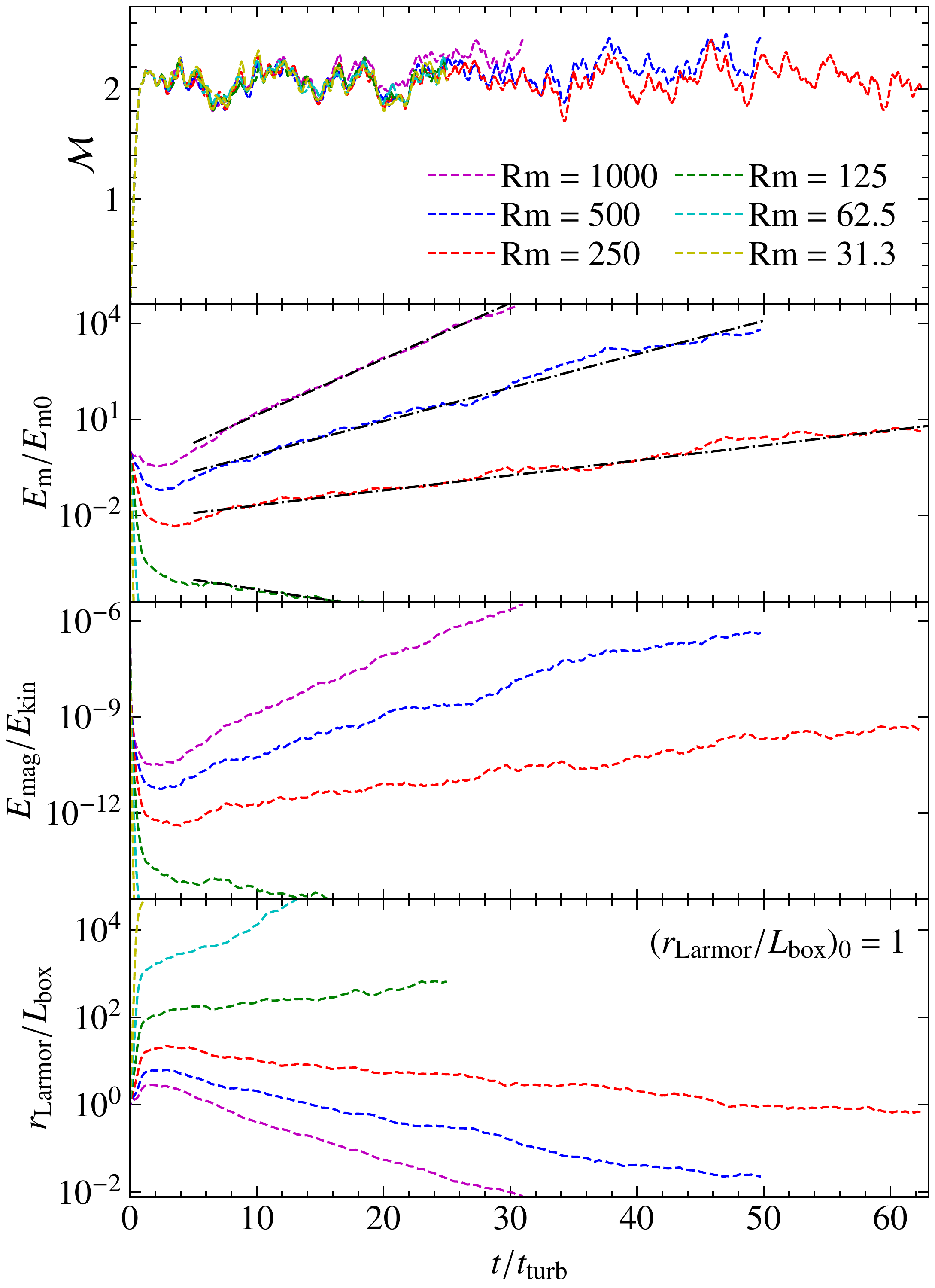} &
\includegraphics[height=0.65\linewidth]{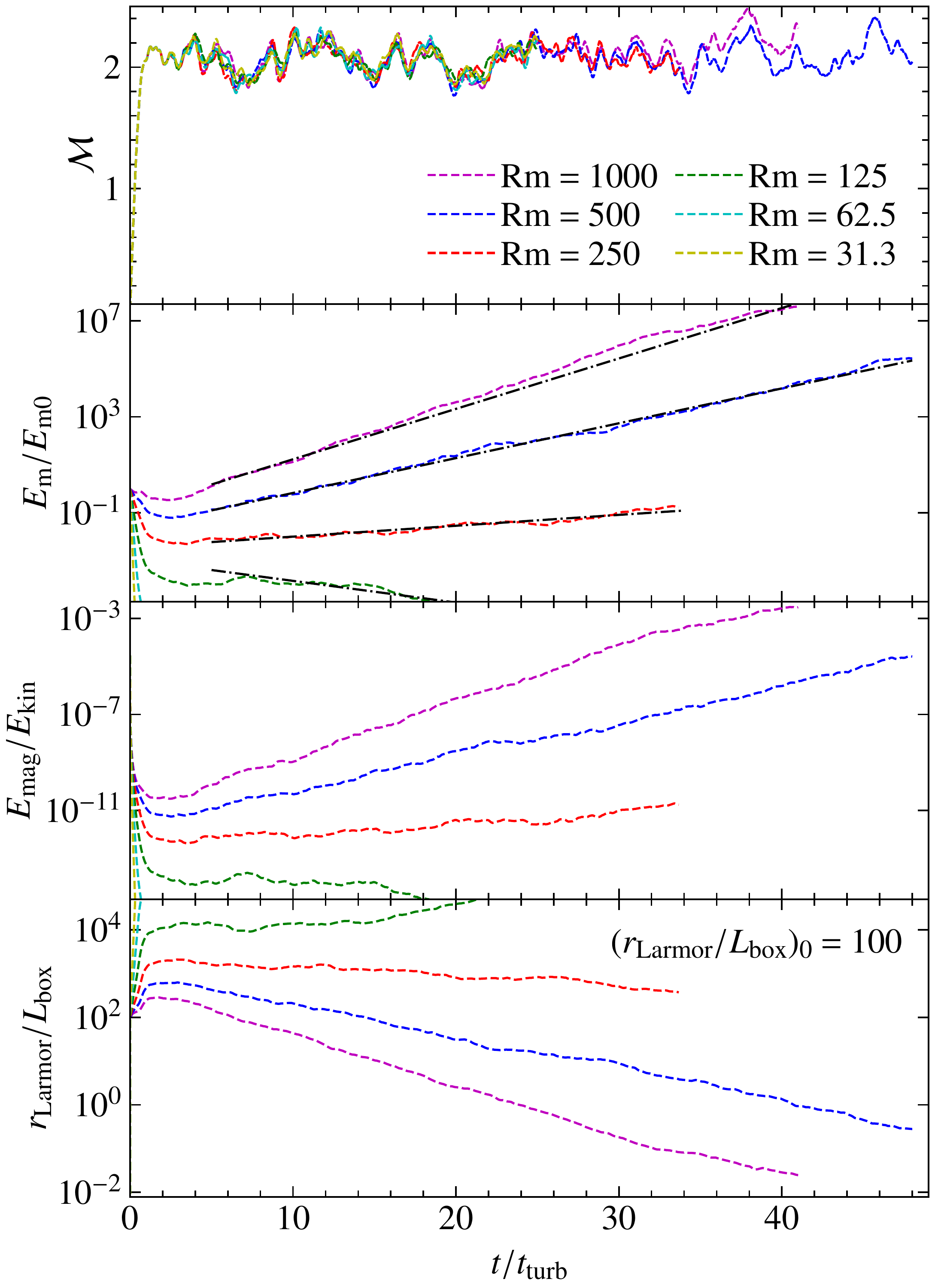} 
\\
\end{tabular}
\end{center}
\caption{Evolution of the Mach number ($\Mach$), magnetic energy normalised by its initial value ($\Emag$), the magnetic-to-kinetic energy ratio ($\ratio$), and the Larmor ratio ($\magnetisation$) as functions of time, expressed in units of the eddy turnover time ($\ted$), for the weakly-collisional HPIC dynamo simulations with magnetic Reynolds numbers $\Rm = 31.3$--1000. The left and right panels show initially magnetised ($\initmagnetisation = 1$) and initially unmagnetised ($\initmagnetisation = 100$) plasmas, respectively. For sufficiently large $\Rm$, the magnetic energy shows exponential amplification, characteristic of the kinematic phase of the turbulent dynamo, i.e., where $\ratio\ll1$ (see third panel). For $\Rm < \Rmcrit$, where $\Rmcrit$ is the critical magnetic Reynolds number, the magnetic field fails to grow and instead decays with time. The black dashed lines in the second panel show the fitted magnetic energy growth or decay rates. The fourth panel shows the evolution of the Larmor ratio, showing that the Larmor radius decreases as the magnetic field is amplified and increases as the magnetic field decays. The measured Mach numbers, together with the dynamo growth/decay rates, are listed in \Tab{table:sims}.}
\label{fig:time_evol}
\end{figure*}

\Fig{fig:time_evol} shows the time evolution for the two initial magnetisation values, $\initmagnetisation = 1$ and $\initmagnetisation = 100$, spanning magnetic Reynolds numbers of $\Rm = 31.3$, 62.5, 125, 250, 500, and 1000. From top to bottom, the four panels show the Mach number, $\Mach$, the magnetic energy normalised by its initial value, $\Emag$, the magnetic-to-kinetic energy ratio, $\ratio$, and the Larmor ratio, $\magnetisation$, respectively. 

The turbulent driving and ion cooling maintain an approximately
constant Mach number, as shown in the top panels of \Fig{fig:time_evol}. After an initial transient phase of about $2\,\ted$, the flow settles into a statistically stationary turbulent state. At sufficiently large $\Rm$, the magnetic energy increases exponentially, marking the kinematic phase of the turbulent dynamo. This phase is the primary focus of this study.

As $\Rm$ is reduced by increasing the Ohmic diffusivity, the dynamo becomes less efficient, leading to lower magnetic energy growth rates. Eventually, below a critical magnetic Reynolds number, $\Rmcrit$, the magnetic field is no longer amplified and instead decays with time. To quantify the dynamo efficiency, we determine the magnetic energy growth rate, $\Gamma$, by fitting the magnetic energy in the kinematic phase with
\begin{equation}
    \Emag = \exp\!\left(\Gamma t\right).
\end{equation}
We exclude the initial transition phase during the first $5\,\ted$
from the fit. The resulting growth and decay curves are overplotted
as black dashed lines in the second panel of \Fig{fig:time_evol}.
Uncertainties in $\Gamma$ are estimated following the method described in appendix~C of \citet{AchikanathChirakkara+2023}, and the fitted values are listed in \Tab{table:sims}.

The third panel of \Fig{fig:time_evol} shows the evolution of the magnetic-to-kinetic energy ratio, $\ratio$. As the magnetic field is amplified by the turbulent dynamo, $\ratio$ increases until the magnetic energy becomes comparable to the kinetic energy of the turbulent eddies at the viscous scale. At this stage, the exponential amplification transitions to a linear growth phase before eventually reaching saturation \citep{Seta&Federrath2020}. We note that the present simulations have not been evolved into the saturation regime, which would require substantially more computational resources and is not required for the measurements of $\Rmcrit$.

The fourth panel of \Fig{fig:time_evol} shows the evolution of the Larmor ratio. As the magnetic field is amplified, the average ion Larmor radius decreases, resulting in stronger plasma magnetisation, whereas magnetic field decay results in an increase in the Larmor radius.

\subsection{Measuring the critical magnetic Reynolds number}
\label{sec:critical_Rm}
\begin{figure*}
    \includegraphics[scale = 0.25]{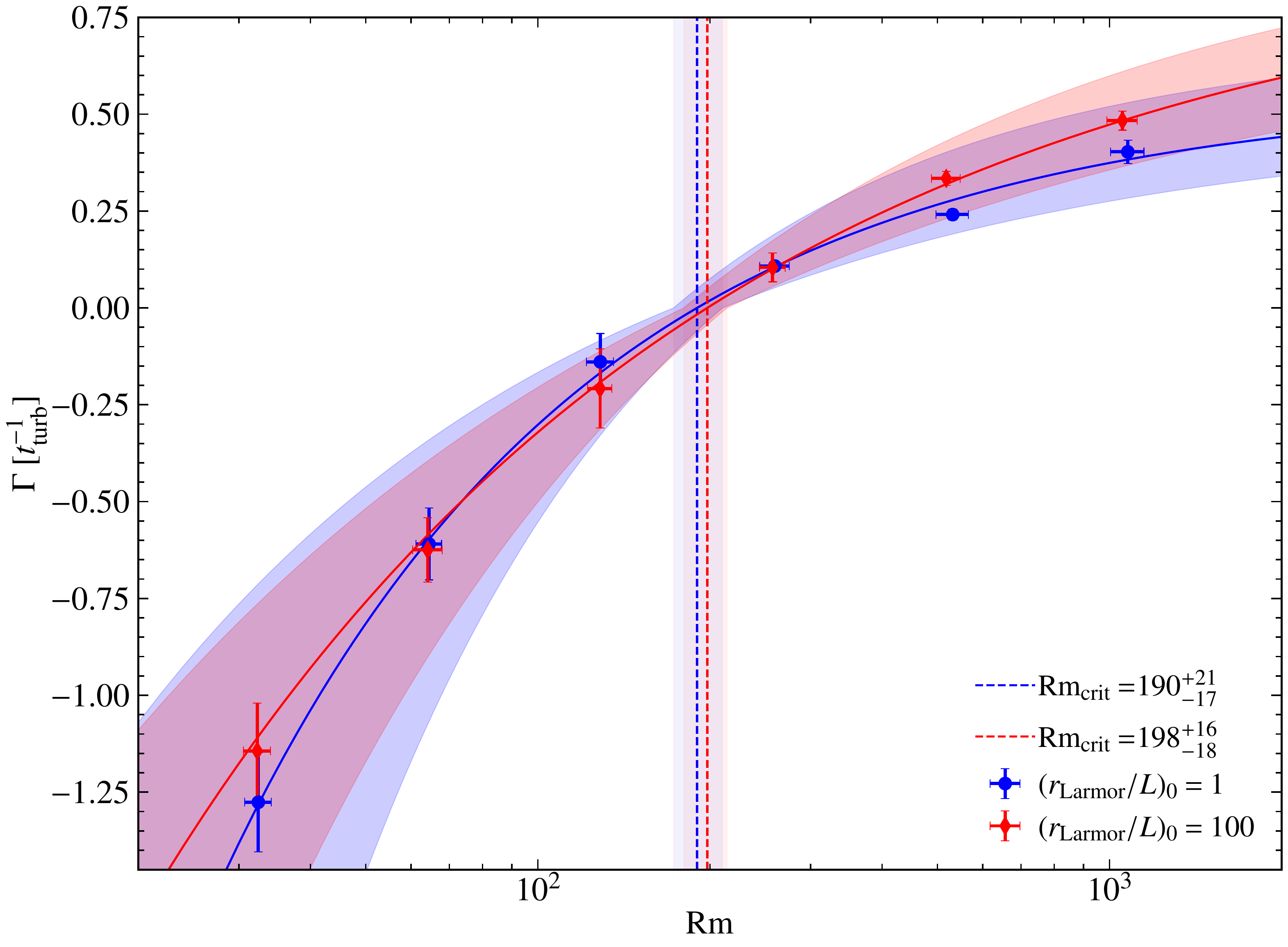}
    \caption{Magnetic energy growth rate ($\Gamma$) as a function of the magnetic Reynolds number ($\Rm$) in the weakly-collisional supersonic dynamo runs with initially magnetised ($\initmagnetisation = 1$) and initially unmagnetised ($\initmagnetisation = 100$) plasmas. The solid curves show the best-fit model given by \Eq{eqn:Rmfit}, and the shaded regions show the uncertainties. The vertical dashed lines mark the fitted values of $\Rmcrit$, which are also listed in the legend. All fit parameters are summarised in \Tab{table:fit_par}.}
    \label{fig:Rmtests_Growthrate}
\end{figure*}

\Fig{fig:Rmtests_Growthrate} shows the energy growth/decay rate $\Gamma$ as a function of $\Rm$ for both initial magnetisation values. For each initial Larmor ratio, we model the dependence of $\Gamma$ on $\Rm$ as
\begin{equation}
    \Gamma (\Rm) = \Gamma_{\rm sat} \left[ 1 - \left( \frac{\Rm}{\Rm_{\rm crit}} \right)^{\alpha} \right],
    \label{eqn:Rmfit}
\end{equation}
where $\Rm_{\rm crit}$, $\Gamma_{\rm sat}$, and $\alpha$ are fit parameters \citep{AchikanathChirakkara+2023}.
In the above equation, $\Rm_{\rm crit}$ is the critical magnetic Reynolds number below which Ohmic diffusivity suppresses dynamo amplification, resulting in magnetic field decay. $\Gamma_{\rm sat}$ represents the asymptotic magnetic energy growth rate in the high-$\Rm$ limit, and $\alpha$ is a power-law exponent that determines how rapidly $\Gamma$ rises above $\Rm_{\rm crit}$ and approaches this asymptotic value. The best-fit curves obtained using \Eq{eqn:Rmfit} are shown in \Fig{fig:Rmtests_Growthrate} as solid lines. The vertical dashed lines show the fitted values of $\Rm_{\rm crit}$, while the shaded regions represent their respective uncertainties. The best-fit values of $\Rm_{\rm crit}$, $\Gamma_{\rm sat}$, and $\alpha$ are listed in \Tab{table:fit_par}. For $\Rm>\Rm_{\rm crit}$, the positive growth rate ($\Gamma>0$) indicates magnetic-field amplification by the weakly collisional dynamo, with $\Rmcrit=190^{+21}_{-17}$ for initially magnetised plasma and $\Rmcrit=198^{+16}_{-18}$ for initially unmagnetised plasma.

\begin{table}
\centering
\caption{Best-fit parameters based on \Eq{eqn:Rmfit}: the critical magnetic Reynolds number ($\Rmcrit$), asymptotic magnetic energy growth rate in the high-$\Rm$ limit ($\Gammasat$), and power-law exponent ($\alpha$), for the two initial Larmor ratios.}
\begin{tabular}{lccc}
\hline
$\initmagnetisation$ & $\Rmcrit$ & $\Gammasat$ & $\expcoeff$\\
\hline
$1$ & $190^{+21}_{-17}$ & $0.55^{+0.14}_{-0.07}$ & $-0.68^{+0.13}_{-0.13}$ \\
$100$ & $198^{+16}_{-18}$ & $0.95^{+0.05}_{-0.17}$ & $-0.43^{+0.03}_{-0.03}$ \\
 \hline
\end{tabular}
\label{table:fit_par}
\end{table}

The critical magnetic Reynolds numbers for the initially magnetised and initially unmagnetised plasmas are the same within their uncertainties, indicating that, in the supersonic regime, $\Rmcrit$ is not affected by the initial plasma magnetisation. This contrasts with our previous study of the weakly-collisional turbulent dynamo in the subsonic regime of turbulence \citep{AchikanathChirakkara+2023}, in which we found that the critical magnetic Reynolds number was marginally larger for initially magnetised plasmas than for initially unmagnetised plasmas.

\section{Comparison with the collisional MHD turbulent dynamo}
\label{sec:comparison}
In this section, we compare the critical magnetic Reynolds number
measured in this work with the $\Rmcrit$ values obtained in collisional MHD studies. \citet{Haugen+2004b} performed direct numerical simulations of compressible, isothermal MHD turbulence to investigate the dependence of $\Rmcrit$ on the Mach number. They estimated $\Rmcrit$ by interpolating between simulations in which the magnetic field grew and those in which magnetic field decayed. They found that $\Rmcrit$ increases sharply around $\Mach \sim 1$, with subsonic and supersonic values of $\Rmcrit \sim 35$ and $70$, respectively, for $\Pm=1$, and $\Rmcrit \sim 25$ and $50$ for $\Pm=5$. These are characteristic values for the two Mach-number regimes rather than measurements at specific Mach numbers. These results show that increasing compressibility raises the threshold for turbulent dynamo action and that $\Rmcrit$ depends on the magnetic Prandtl number. Using simulations of highly compressible MHD turbulence with Mach numbers in the range
$4\lesssim\Mach\lesssim11$, \citet{Federrathetal2014ApJ} studied turbulent dynamo action over the range $0.1\leq\Pm\leq10$ and found a critical magnetic Reynolds number of $\Rmcrit=129^{+43}_{-31}$ in the supersonic regime. This further supports the conclusion that the higher the compressibility (through increased $\Mach$), the higher $\Rmcrit$.

For our $\Mach=2$ HPIC experiments of weakly-collisional plasmas, we find $\Rmcrit = 190^{+21}_{-17}$ and $198^{+16}_{-18}$ in the initially magnetised and unmagnetised case, respectively. Thus, $\Rmcrit$ in supersonic, weakly-collisional plasmas is higher than in collisional MHD plasmas. We note that, for a more direct comparison with MHD, we require the magnetic Prandtl number. However, neither $\Re$ nor $\Pm$ is a fixed parameter in HPIC simulations, making this comparison difficult. Both $\Re$ and $\Pm$ are self-consistently determined by wave-particle interactions in a weakly-collisional plasma and can evolve as the magnetic field strength changes due to dynamo action.

Our results further reveal that the value of $\Rmcrit$ for the weakly-collisional turbulent dynamo is higher in the supersonic regime than in the subsonic regime \citep{AchikanathChirakkara+2023}. This is consistent with the trend
seen in MHD simulations. For the subsonic weakly-collisional turbulent dynamo, \citet{AchikanathChirakkara+2023} found that an initially magnetised plasma had a marginally higher value of
$\Rmcrit\approx124\pm8$ than an initially unmagnetised plasma, for
which $\Rmcrit\approx107\pm4$. We find that this is not the case in
the supersonic regime.

Using a Wentzel--Kramers--Brillouin (WKB) approximation to the
\citet{Kazantsev1968} equation, \citet{Schoberetal2012} found
$\Rmcrit\approx110$ for Kolmogorov turbulence and
$\Rmcrit\approx2700$ for Burgers turbulence, showing that increasing
compressibility makes the dynamo instability more difficult to
excite. \citet{Bovino+2013} investigated the onset of the turbulent
dynamo by numerically solving the Kazantsev equation
in the regimes $\Pm\ll1$ and $\Pm\gg1$, for both incompressible Kolmogorov and highly-compressible Burgers turbulence. They predicted that the critical magnetic Reynolds number depends strongly on the nature of the turbulence, increasing from $\Rmcrit\approx320$ for Kolmogorov turbulence to $\Rmcrit\approx3.2\times10^4$ for Burgers turbulence, again implying that highly compressible turbulence substantially suppresses turbulent dynamo action. We note that the approach used in these theoretical studies differs from the 3-D HPIC simulations used here. As in simulations of the supersonic MHD turbulent dynamo, the weakly-collisional dynamo in the supersonic regime exhibits critical magnetic Reynolds numbers that are substantially lower than those predicted by these semi-analytical studies, which deserves further investigation into the theoretical modelling of turbulent dynamo amplification in both collisional and weakly-collisional plasmas.

\section{Conclusions}
\label{sec:conclusions}
Using three-dimensional hybrid particle-in-cell (HPIC) simulations,
we determined the critical magnetic Reynolds number of the
weakly-collisional turbulent dynamo in the supersonic regime. We drove purely solenoidal turbulence at $\Mach\approx2$ and measured the magnetic Reynolds number over the range $\Rm\approx31$--$1000$. We considered initially magnetised and initially unmagnetised plasmas
with initial Larmor ratios $\initmagnetisation=1$ and $100$, respectively.

We find that in the kinematic regime of the dynamo, the density and velocity structures show little variation with $\Rm$, whereas the magnetic field becomes progressively smoother as $\Rm$ decreases because of increasing Ohmic diffusion. At sufficiently large $\Rm$, the magnetic energy grows exponentially during the kinematic phase of the dynamo, while at sufficiently low $\Rm$, magnetic diffusion dominates and the magnetic energy decays. By fitting the magnetic-energy growth rate as a function of $\Rm$, we obtain $\Rmcrit=190^{+21}_{-17}$ and $198^{+16}_{-18}$ for the initially magnetised and unmagnetised case, respectively.

The two measurements of $\Rmcrit$ are the same within their
uncertainties, indicating that the initial plasma magnetisation does
not significantly affect the onset of the weakly-collisional
turbulent dynamo in the supersonic regime. This differs from the
marginal dependence on initial magnetisation found previously for the
weakly-collisional dynamo in the subsonic regime. The values of
$\Rmcrit$ in the supersonic regime are also higher than the corresponding subsonic
measurements, indicating that increased compressibility makes the
weakly-collisional dynamo more difficult to excite. Consistent with collisional MHD simulations, compressibility raises the threshold for dynamo action and our measured values, $\Rmcrit\simeq190$--$198$, are approximately 1.5 times higher than the supersonic MHD value $\Rmcrit=129^{+43}_{-31}$ found by \citet{Federrathetal2014ApJ}.

\section*{Acknowledgements}
C.F.~acknowledges funding provided by the Australian Research Council (Discovery Projects DP230102280 and DP250101526), and the Australia-Germany Joint Research Cooperation Scheme (UA-DAAD). This research was enabled in part by support provided by Compute Ontario (computeontario.ca) and the Digital Research Alliance of Canada (alliancecan.ca). Computations were performed on the Trillium supercomputer at the SciNet HPC Consortium. SciNet is funded by Innovation, Science and Economic Development Canada; the Digital Research Alliance of Canada; the Ontario Research Fund: Research Excellence; and the University of Toronto \citep{Scinet1, Scinet2}. We further acknowledge high-performance computing resources provided by the Leibniz Rechenzentrum and the Gauss Centre for Supercomputing (grant~pr32lo), the Australian National Computational Infrastructure (grant~ek9) and the Pawsey Supercomputing Centre (project~pawsey0810) in the framework of the National Computational Merit Allocation Scheme and the ANU Merit Allocation Scheme. The simulation software \flash, which forms the bases for \ahkash, was in part developed by the DOE-supported Flash Center for Computational Science at the University of Chicago and the University of Rochester. All revisions were reviewed by the authors.

\section*{Data Availability}
The simulation data will be shared upon reasonable request to Radhika Achikanath Chirakkara, noting that the simulations have a total size of 17~TB.



\bibliographystyle{mnras}
\bibliography{Ref_files_Radhika.bib}



\appendix
\section{Density, velocity, and magnetic field structures for initially unmagnetised plasma}
\label{sec:plots_rL100}

\begin{figure*}
    \centering
    \includegraphics[scale=0.9]{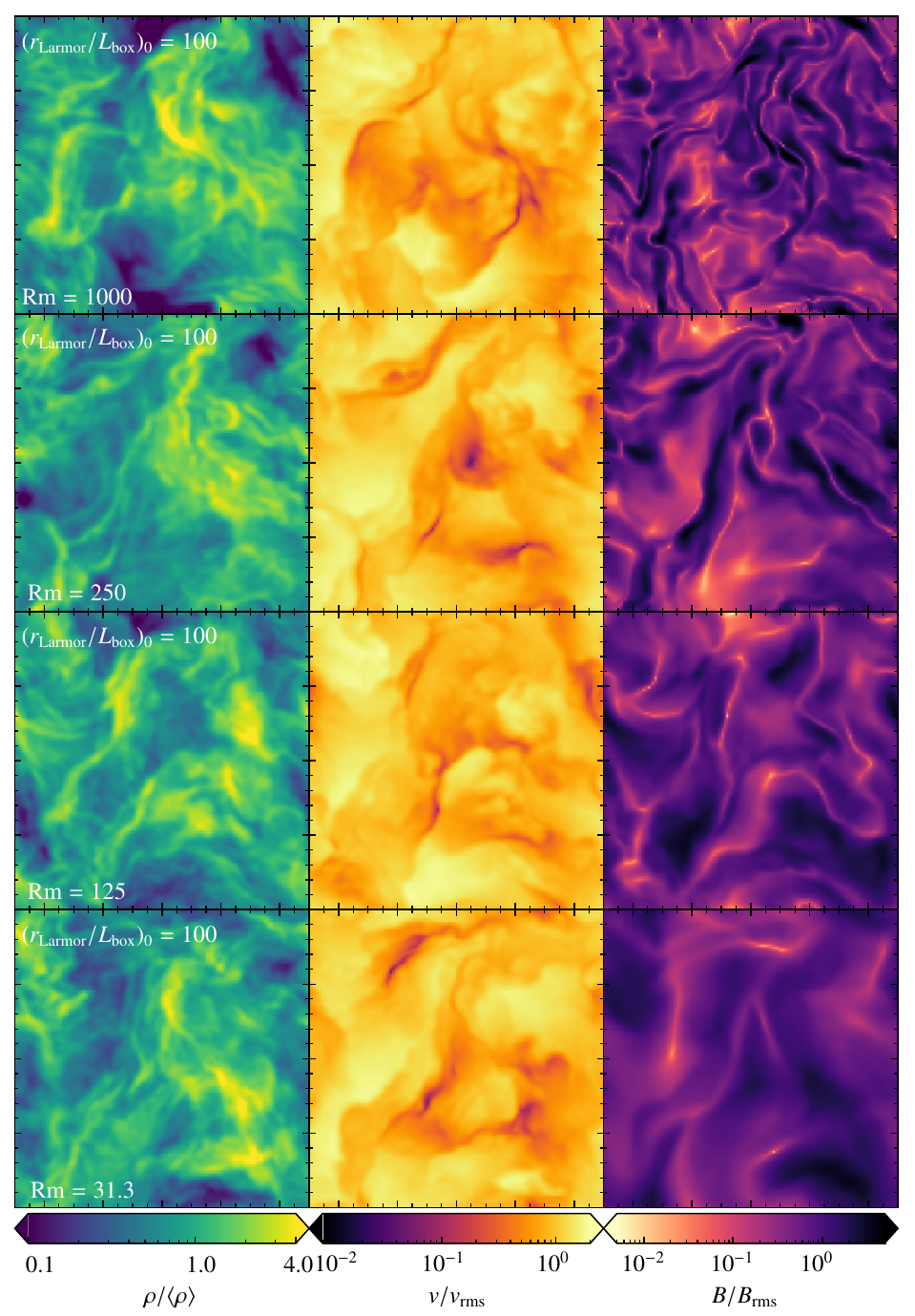}
    \caption{Same as \Fig{fig:slice_plots_rL1}, but for initially unmagnetised simulations with $\initmagnetisation = 100$.}
    \label{fig:slice_plots_rL100}
\end{figure*}

\Fig{fig:slice_plots_rL100} shows the density, velocity, and magnetic-field structure for the supersonic turbulent dynamo simulations with an initial Larmor ratio of $\initmagnetisation = 100$. The overall behaviour is qualitatively similar to that discussed in \Sec{sec:slice_plots} for the initially magnetised runs. The density and velocity fields exhibit little dependence on the magnetic Reynolds number, while the magnetic field becomes progressively smoother as $\Rm$ decreases.

\bsp	
\label{lastpage}
\end{document}